\documentclass[10pt]{article}
\usepackage{graphicx,amssymb}
\newcommand{\be}{\begin{equation}}
\newcommand{\ee}{\end{equation}}
\newcommand{\beq}{\begin{eqnarray}}
\newcommand{\eeq}{\end{eqnarray}}

\begin{document}

\title{On the Comment by Palle (astro-ph/0503562) on the vorticity and shear of the Universe}

\author{ Sigbj{\o}rn  Hervik\\
{\it  \small Dalhousie University, Dept. of Mathematics and Statistics,}\\
{\it  \small Halifax, NS, Canada B3H 3J5}\\
{\tt\small herviks@mathstat.dal.ca}\\}
\date{}
\maketitle
\begin{abstract}
We show that a recent comment where Palle criticises works regarding CMBR fluctuations for a particular type of Bianchi models, is incorrect. In particular, in contrast to the claims of Palle, we point out that Bianchi models do allow for vorticity.  
\end{abstract}
In a recent work Jaffe \emph{et al} \cite{Jaffe} attempt to fit WMAP data with CMBR calculations for some spatially homogenenous Bianchi models.  Soon after this paper appeared, a comment by Palle \cite{Palle} criticises this work as well as an older paper by Barrow \emph{et al} \cite{BJS}. Part of Palle's criticism is based on an erroneous claim regarding the possibility of vorticity for these Bianchi models. 

For the spatially homogeneous Bianchi models there are two naturally defined time-like unit vector fields: (\textit{i}) the unit vector field, $u^{\mu}$, normal to the group orbits, and (\textit{ii}) the four-velocity, $\hat{u}^{\mu}$, of the perfect fluid. The vector field $u^{\mu}$ describes the evolution of the spatially homogeneous hypersurfaces and in the study of Bianchi models this vector field is usually taken to be the four-velocity of the fundamental observers \cite{DynSys,COLEY}. This vector field necessarily has zero vorticity, but in general gives rise to a non-zero shear tensor $\sigma_{ab}$. Regarding the fluid four-velocity $\hat{u}^{\mu}$, this may or may not be aligned with $u^{\mu}$. If $\hat{u}^{\mu}$ is not aligned with $u^{\mu}$ we call the model \emph{tilted} and non-tilted otherwise. It is clear that the fluid can only have vorticity if the fluid is tilted (when we refer to vorticity in a cosmological model we usually refer to the fluid vorticity!). 

In \cite{Palle} Palle erroneously claims that \textit{"..it is very well known that standardly defined spacetime vorticity vanishes for these Bianchi models"} and then refers to \cite{EM} which only considers non-tilted models (hence, zero vorticity by assumption). A more appropriate reference is \cite{KingEllis} which considers tilted Bianchi models. This reference also discusses vorticity and gives conditions for when a tilted Bianchi model has vorticity.  Clearly, based on these conditions Bianchi models can have vorticity and, in fact, non-zero vorticity is the general case. In particular, Bianchi type VII$_h$ models can have two non-zero vorticity components as pointed out by Barrow \textit{et al} \cite{BJS}. Other papers that consider vortical motion in Bianchi cosmologies are, for example, \cite{CH,Lukash}. 

Palle also states that the vorticity of \cite{BJS} is \emph{"just certain space rotation proportional to the shear components"}. In principle, the vorticity of the fluid and the shear of the hypersurfaces are independent quantities. However, in cosmology we also impose the Einstein field equation which relates the stress energy tensor of the matter to the Ricci curvature of the universe. In the orthonormal frame formalism, the Einstein field equation imposes constaints which relate some of the matter variables with the variables for the geometry \cite{DynSys,COLEY}. In eqs. (4.3) and (4.4) of ref \cite{BJS}, two of these constraints have been solved to relate the vorticity of the matter to the shear of the spacetime for which it induces. Hence, Palle's "space rotation" is in fact some of the components of the Einstein field equation that has been solved to obtain expressions for the vorticity.  

Moreover, Palle expresses concern regarding the gauge invariance of some relations regarding the shear. In the background, which both \cite{Jaffe} and \cite{BJS} assume to be FRW, the shear vanishes, and hence, when perturbed, the shear will be gauge invariant \cite{SW}. Thus the corresponding expressions for the CMBR will also be gauge invariant. 

Based on this we can see that the criticism by Palle is based on claims that are erroneous, or at best, misleading.   

\bibliographystyle{amsplain}

\end{document}